\documentclass[3p,times,twocolumn]{elsarticle}

\usepackage{ecrc}

\volume{00}

\firstpage{1}

\journalname{Nuclear Physics B Proceedings Supplement}

\runauth{W.~Gradl}

\jid{nuphbp}

\jnltitlelogo{Nuclear Physics B Proceedings Supplement}

\usepackage{amsmath}
\usepackage{amssymb}

\biboptions{sort&compress}

\usepackage[figuresright]{rotating}
\usepackage{relsize}
\usepackage[utf8]{inputenc}
\usepackage{xspace}
\usepackage{tikz}
\usepackage{booktabs}
\usepackage{hyperref}

\definecolor{darkgreen}{rgb}{0.05,0.5,0}

\def\calB  {{\ensuremath{\cal B}}\xspace}
\def\jpsi  {\ensuremath{{J\mskip -3mu/\mskip -2mu\psi\mskip 2mu}}\xspace}
\def\invfb {\ensuremath{\mbox{\,fb}^{-1}}\xspace}
\def\Kbar  {\kern 0.2em\overline{\kern -0.2em K}{}\xspace}
\def\Kp    {\ensuremath{K^+}\xspace}
\def\Km    {\ensuremath{K^-}\xspace}
\def\Kz    {\ensuremath{K^0}\xspace}
\def\Kzb   {\ensuremath{\Kbar^0}\xspace}
\def\KzKzb {\ensuremath{\Kz \kern -0.16em \Kzb}\xspace}
\def\KS    {\ensuremath{K^0_{\scriptscriptstyle S}}\xspace} 
\def\KL    {\ensuremath{K^0_{\scriptscriptstyle L}}\xspace} 
\def\Kstar {\ensuremath{K^*}\xspace}
\def\Kstarb{\ensuremath{\Kbar^*}\xspace}
\def\piz   {\ensuremath{\pi^0}\xspace}
\def\pip   {\ensuremath{\pi^+}\xspace}
\def\pim   {\ensuremath{\pi^-}\xspace}
\def\epem  {\ensuremath{e^+e^-}\xspace}
\def\gammaisr{\ensuremath{\gamma_\text{ISR}}\xspace}
\def\amu   {\ensuremath{a_\mu}\xspace}
\newcommand{\kev}{\ensuremath{\mathrm{\,ke\kern -0.1em V}}\xspace}
\newcommand{\mev}{\ensuremath{\mathrm{\,Me\kern -0.1em V}}\xspace}
\newcommand{\mevcc}{\ensuremath{\mathrm{\,Me\kern -0.1em V}/c^2}\xspace}
\newcommand{\gev}{\ensuremath{\mathrm{\,Ge\kern -0.1em V}}\xspace}
\newcommand{\gevcc}{\ensuremath{\mathrm{\,Ge\kern -0.1em V}/c^2}\xspace}
\mathchardef\Upsilon="7107
\def\Y#1S{\ensuremath{\Upsilon{(#1S)}}\xspace}
\def\FourS {\Y4S}

\newcommand{\babar}{\mbox{\slshape B\kern-0.01em{\smaller A}\kern-0.01em
    B\kern-0.05em{\smaller A\kern-0.05em R}}\xspace}

\newcommand{\KSKLpiz}{\ensuremath{\KS\KL\piz}\xspace}
\newcommand{\KSKLpizpiz}{\ensuremath{\KS\KL\piz\piz}\xspace}

\begin{document}

\begin{frontmatter}



\dochead{}

\title{New ISR Cross Section Results on \KSKLpiz and \KSKLpizpiz From \protect\babar{} }


\author{Wolfgang Gradl}
\ead{gradl@uni-mainz.de}
\address{Institute for Nuclear Physics, Johannes Gutenberg University,
  D-55099 Mainz}

\author{representing the \babar collaboration}

\begin{abstract}
We present preliminary measurements of the cross sections for $\epem \to \KSKLpiz$
and $\KSKLpizpiz$ obtained using the technique of Initial State
Radiation with $469\invfb$ of $\epem$ collision data collected  with
the \babar detector at or near the $\FourS$ resonance.  The resonant
substructure of $\KSKLpiz$ is investigated, and branching fractions
for the decays of the charmonium resonances $\jpsi$ and $\psi(2S)$
into these final states are measured.
\end{abstract}


\end{frontmatter}


\section{Introduction}
\label{sec:introduction}

The anomalous magnetic moment of the electron $a_e \equiv(g_e -2) / 2$
presents an extremely precise test of Quantum Electrodynamics (QED);
its experimental value and theoretical calculations agree to about one
part in a billion.  In contrast, there has been a long-standing
tension between the standard-model (SM) expectation
\cite{Davier:2010nc,Hagiwara:2011af} and the experimental
measurement~\cite{Bennett:2006fi} for the anomalous magnetic moment of
the muon $\amu$.  More specifically, the experimentally determined value
$10^{10}\times a_\mu^\text{exp} = 11659208.9 \pm 6.3$ and the theory
value $10^{10}\times a_\mu^\text{SM} = 11659180.2\pm4.9$
\cite{Davier:2010nc} differ by $10^{10} \times \Delta a_\mu = 28.7 \pm
8.0$, a discrepancy of about $3.6 \sigma$.
New experiments planned at Fermilab \cite{Grange:2015eea}
and J-PARC \cite{Otani:2015jra} are set to decrease the experimental
uncertainty by factors of four or better.

The discrepancy in $\amu$ may be due to physics beyond the standard
model, but it may also be due to the QCD contributions within the SM,
in particular the hadronic vacuum polarisation (HVP) and
hadronic light-by-light scattering (hLbL) (see
\emph{e.g.}~\cite{Redmer:Tau2016} and references therein).  Due to the
non-perturbative nature of QCD at these energy scales, an \emph{ab
  initio} calculation of the hadronic contributions to \amu is
impossible with current methods.  However, the HVP contributions can
be reliably determined using a dispersion relation and the optical
theorem connecting the HVP to the total cross section for
$\epem \to \text{hadrons}$ at low energies.  Using slightly different
approaches, the HVP contribution is estimated to be
$10^{10} \times \amu^\text{had,LO} = 692.3 \pm 4.2$
\cite{Davier:2010nc} or
$10^{10} \times a_\mu^\text{had, LO} = 694.9 \pm 4.3$
\cite{Hagiwara:2011af}.  The next-to-leading order contribution is
estimated to be
$10^{10}\times a_\mu^\text{had, NLO} = -9.84 \pm 0.06 \pm 0.04$
\cite{Hagiwara:2011af}.

One of the limiting factors in the determination of the HVP is the
need to extrapolate the measured cross section for $\Kp\Km\piz(\piz)$ to
the process $\KzKzb\piz(\piz)$.  The new measurements presented in
this talk will obviate the need for this isospin extrapolation and
should help to decrease the uncertainty on the SM expectation for
\amu.  This measurement is part of an extensive program at \babar to
measure exclusive cross sections of $\epem \to \text{hadrons}$ for a
wide range of final states, at center-of-mass energies between
threshold and $4.5\gev$ (see \emph{e.g.} \cite{Griessinger:Tau2016} for
new measurements for the final states $\pip\pim\piz\piz$ and $\pip\pim\eta$).

\section{Experimental Setup, Event Selection and Reconstruction}
The \babar detector was located at the PEP-II asymmetric-energy
$\epem$ collider at SLAC.  Its construction and performance is
described in great detail in
Refs.~\cite{Aubert:2001tu,TheBABAR:2013jta}.  The analysis presented
here uses 469 \invfb of $\epem$ collision data taken at or just below
the $\Upsilon(4S)$ resonance.  The effective center-of-mass energy
$\sqrt{s^\prime}$ of the collision is lowered by the radiation of a
hard photon from one of the initial-state leptons.  This technique,
which is made possible by the very high luminosity of the \babar dataset,
allows for the simultaneous measurement of hadronic cross sections over a
large range of $\sqrt{s^\prime}$, reducing the point-to-point
systematic uncertainties compared to direct scan experiments.

In the analysis presented here, the reconstruction of events proceeds
through the selection of an ISR photon $\gammaisr$ with an energy of at
least $3\gev$ in the \epem center-of-mass (CM) frame.  The neutral pions and kaons are
reconstructed in their decays $\piz\to\gamma\gamma$ and
$\KS\to\pip\pim$, respectively.  The long-lived, neutral \KL cannot be
fully reconstructed.  Therefore, this analysis uses \KL interacting in the
electromagnetic calorimeter; the direction of the cluster is used, but
the momentum of the \KL is determined by a kinematic fit \cite{Ivanchenko:KinFit,Aulchenko:1998xy} imposing
overall energy-momentum conservation with additional mass
constraints for the $\piz\to\gamma \gamma$ candidate(s).
The resolution in the $\KS\KL\piz(\piz)$ invariant mass is
about $25\mevcc$.

\section{Results}

For both processes under consideration, the reconstruction efficiency
is estimated using Monte Carlo (MC) simulated data.  These MC derived
efficiencies are then corrected to match efficiencies measured on real
data in a range of control channels.  Backgrounds from known ISR
processes seem only to describe about half the background visible in
data; these additional backgrounds are subtracted using events in the
$\chi^2$ control region.

\subsection{\KSKLpiz Cross Section}
\begin{figure}
  \centering
      \begin{tikzpicture}
        \node[anchor=south west]{\includegraphics[width=0.9\columnwidth]{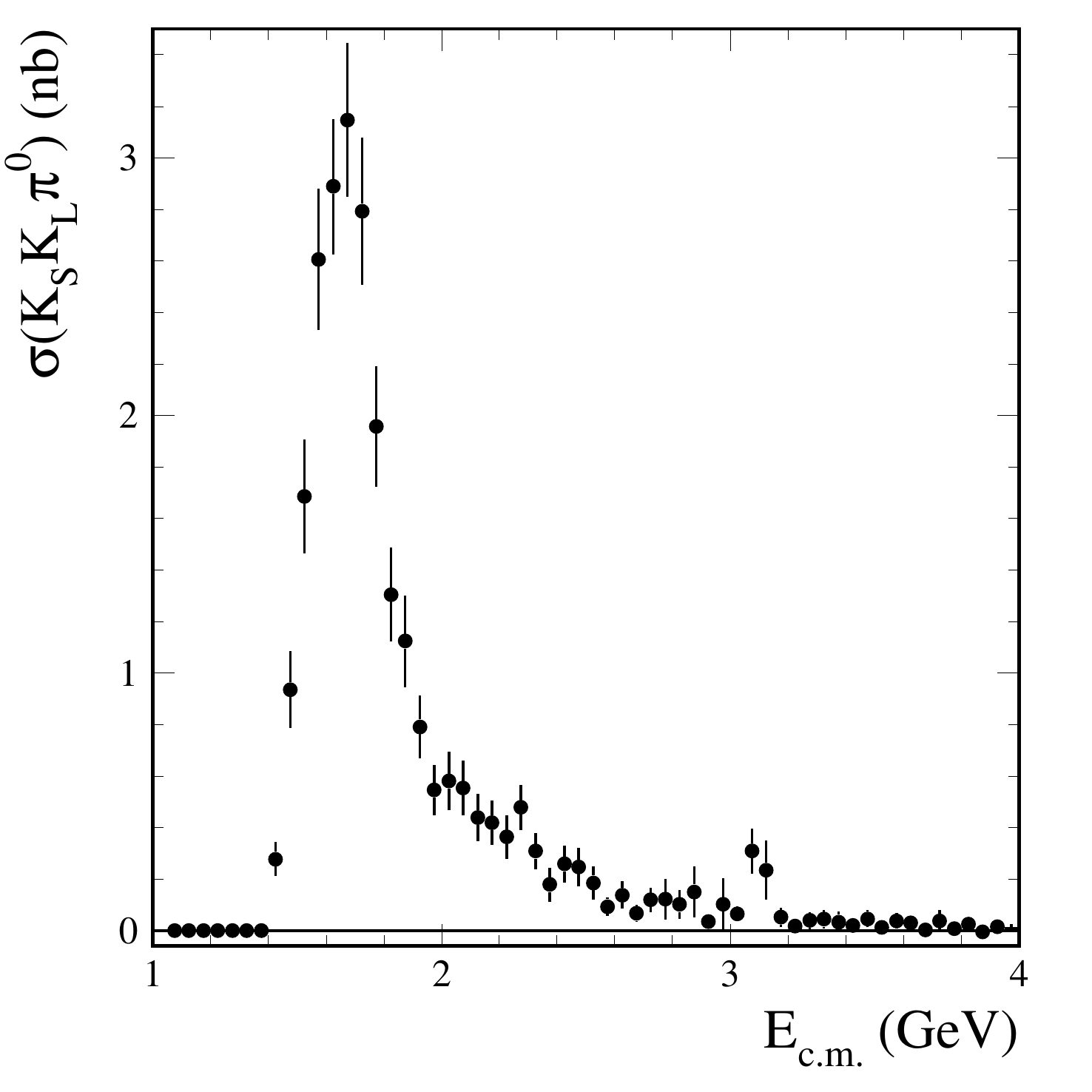}};
    \node[darkgreen,align=center] at (5.0,6) {\babar\\ preliminary \\
      \small stat. uncertainties only};
    \end{tikzpicture}
  
  \caption{Measured cross section for $\epem \to \KSKLpiz$.  The
    uncertainties shown are statistical only. }
  \label{fig:KSKLpiz-xsect}
\end{figure}
 
Events with $\chi^2(\KSKLpiz)<25$ of the kinematic fit are
selected for further analysis, while events in the control region
$25<\chi^2(\KSKLpiz)<50$ are used to determine the background.
After subtracting the background and correcting for the reconstruction
efficiency, the measured cross section for $\epem \to \KSKLpiz$ is
obtained as shown in Fig.~\ref{fig:KSKLpiz-xsect}, where the error bars indicate the
statistical uncertainty only.  No previous measurements of this
process have been reported.  The cross section is dominated by a
broad structure around $1.7\gevcc$, and a small contribution from the
process $\jpsi \to \KSKLpiz$ is visible.  

Systematic uncertainties arise mainly 
due to the uncertainty in the background yields. The corrections
applied to the reconstruction efficiency determined from simulation
contribute a systematic uncertainty of $1.6\%$.  For \KSKLpiz
invariant masses below $2.2\gevcc$, the systematic uncertainty due to
the background subtraction is below $10\%$, rising to $80$--$100\%$
above $3.2\gevcc$, where the event yield is dominated by background
processes.  Radiative corrections are within one percent of unity,
with an overall uncertainty of $1\%$.

\subsection{\KSKLpiz Resonant Substructure}

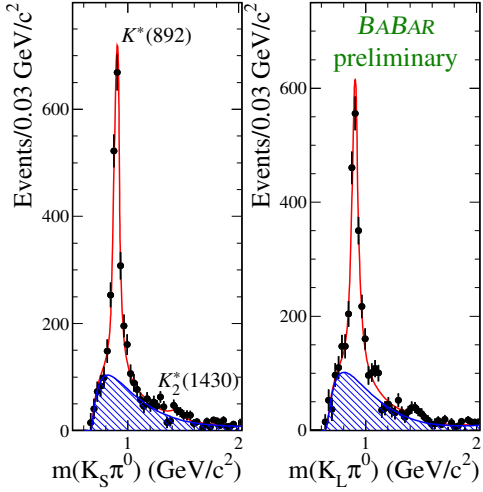
\begin{figure}
  \centering
  \begin{tikzpicture}
    \node[anchor=south west]{\includegraphics[width=0.8\columnwidth]{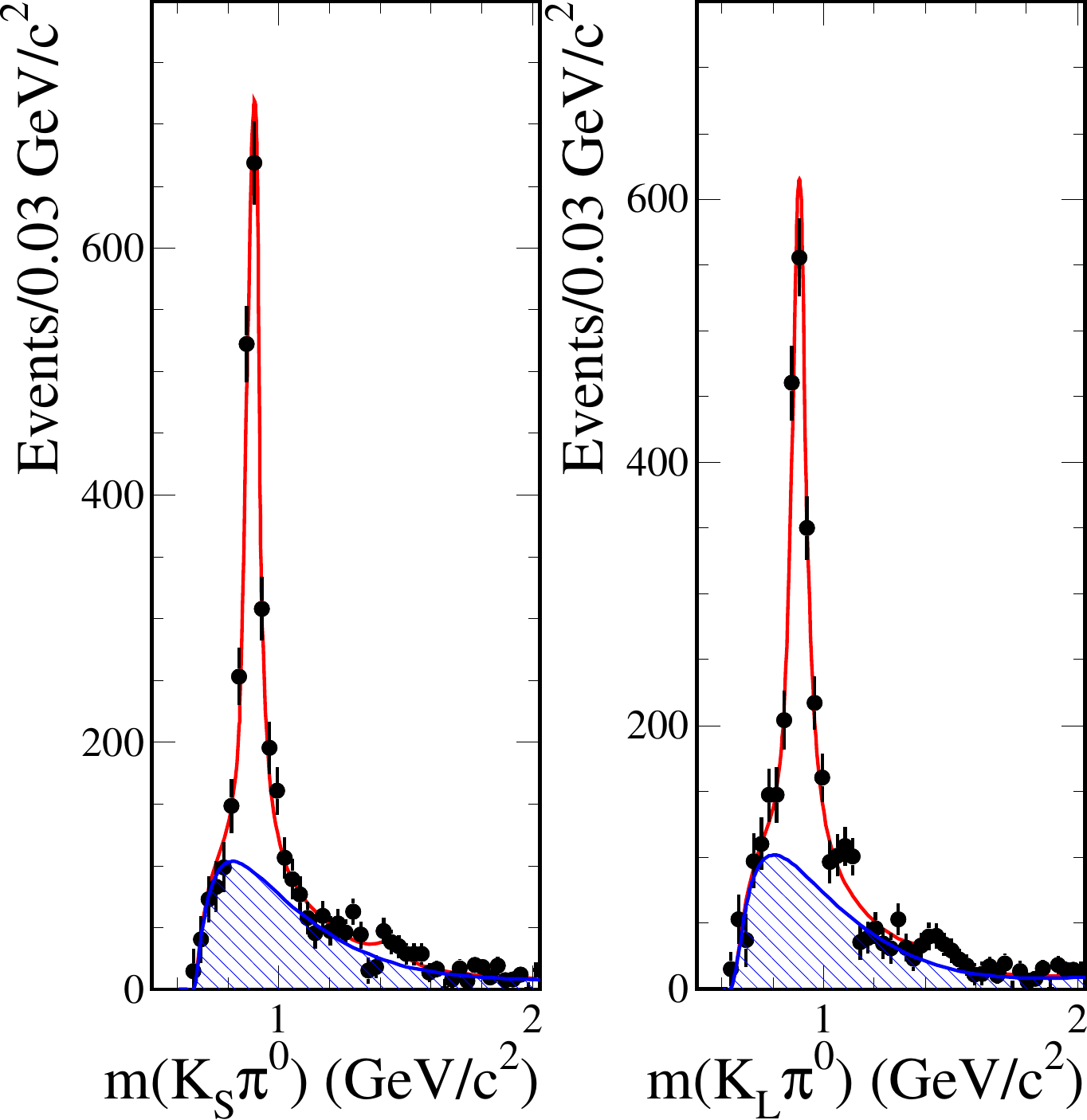}};
    \node[darkgreen,align=center] at (5.25,6) {\babar\\ preliminary};
    \node[align=center] at (2.1,6.1) {\smaller $\Kstar(892)$};
    \node[align=center] at (2.65,1.5) {\smaller $\Kstar_2(1430)$};
  \end{tikzpicture}  
  \caption{Invariant masses of the $\KS\piz$ and $\KL\piz$ subsystems
    in $\epem \to \KSKLpiz$.  Black points represent
    background-subtracted data, the (red) curve shows the result of
    the fit described in the text, and the hatched area shows the
    non-resonant component.}
  \label{fig:KSKLpiz-Kstars}
\end{figure}

A study of the $\KS\piz$ and $\KL\piz$ invariant mass distributions
reveals that the cross section is dominated by the process $\epem \to
\Kstar(892)^0 \Kzb + c.c.$, with a small contribution from
$\Kstar_2(1430) \Kzb + c.c.$ (Fig.~\ref{fig:KSKLpiz-Kstars}).  
We fit the mass spectra with a sum of two Breit-Wigner functions  together with a
function describing the non-resonant contribution.  On
closer inspection, we find that the process $\Kstar(892)^0 \Kzb + c.c.$ almost
saturates the cross section.

\begin{figure}
  \centering
  \begin{tikzpicture}
    \node[anchor=south west]{\includegraphics[width=0.9\columnwidth]{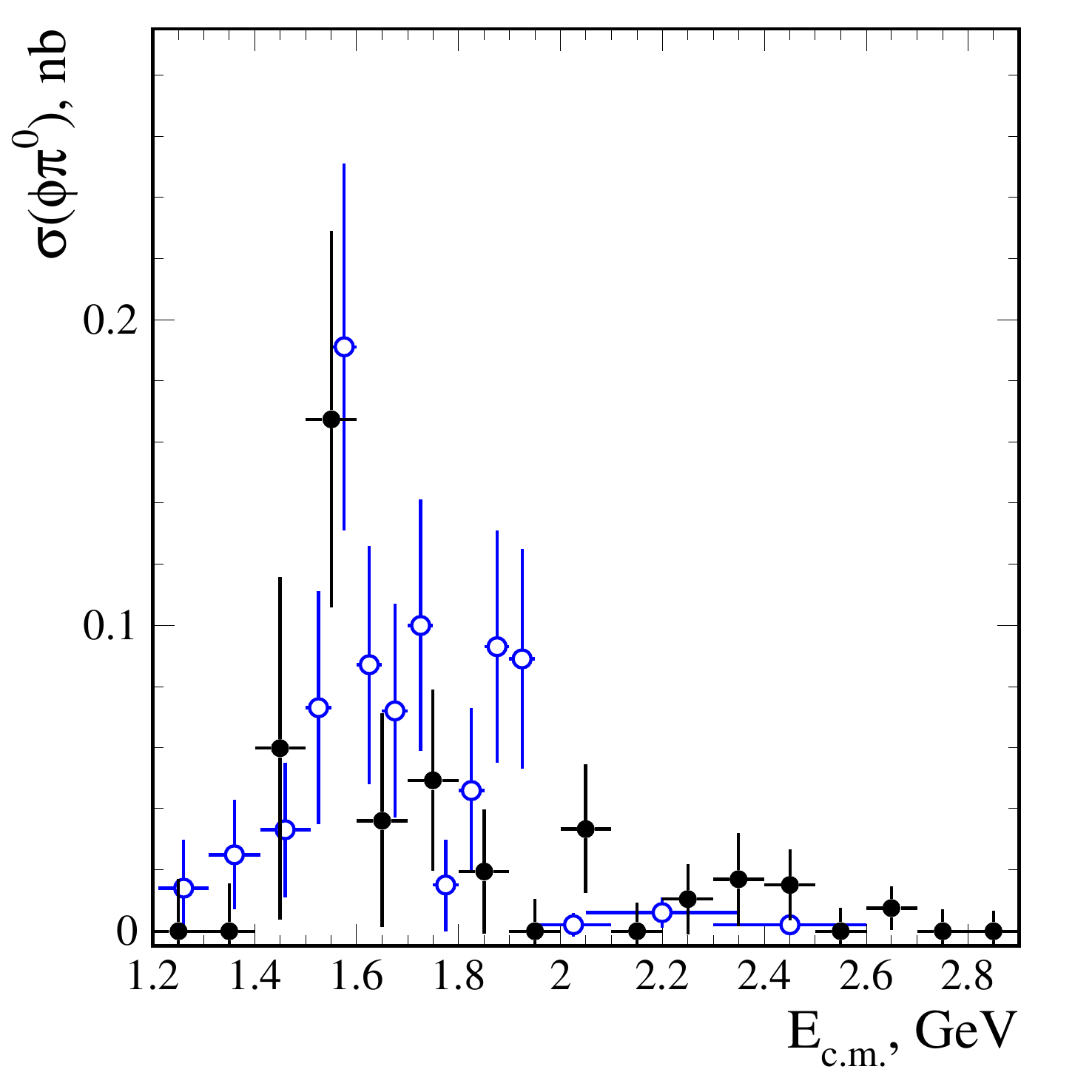}};
    \node[darkgreen,align=center] at (5.0,6) {\babar\\ preliminary};
  \end{tikzpicture}
  \caption{Cross section for $\epem\to\phi\piz$.  The black filled
    points show this measurement, the blue open points show the cross
    section obtained in the final state
    $\Kp\Km\piz$~\cite{Aubert:2007ym}.   In both final states, a
    possible resonant structure near 1.6\gevcc is clearly visible.}
  \label{fig:phipiz}
\end{figure}

A small contribution to the cross section comes
from the isospin $I=1$, OZI suppressed process $\epem \to \phi \piz \to \KSKLpiz$ (see
Fig.~\ref{fig:phipiz}).  The measured cross section is compatible with
the one measured in $\Kp\Km\piz$ \cite{Aubert:2007ym} and shows a
similar peaking structure around $1.6\gevcc$.

\subsection{\KSKLpizpiz Cross Section}

\begin{figure}
  \centering
      \begin{tikzpicture}
        \node[anchor=south west]{\includegraphics[width=0.9\columnwidth]{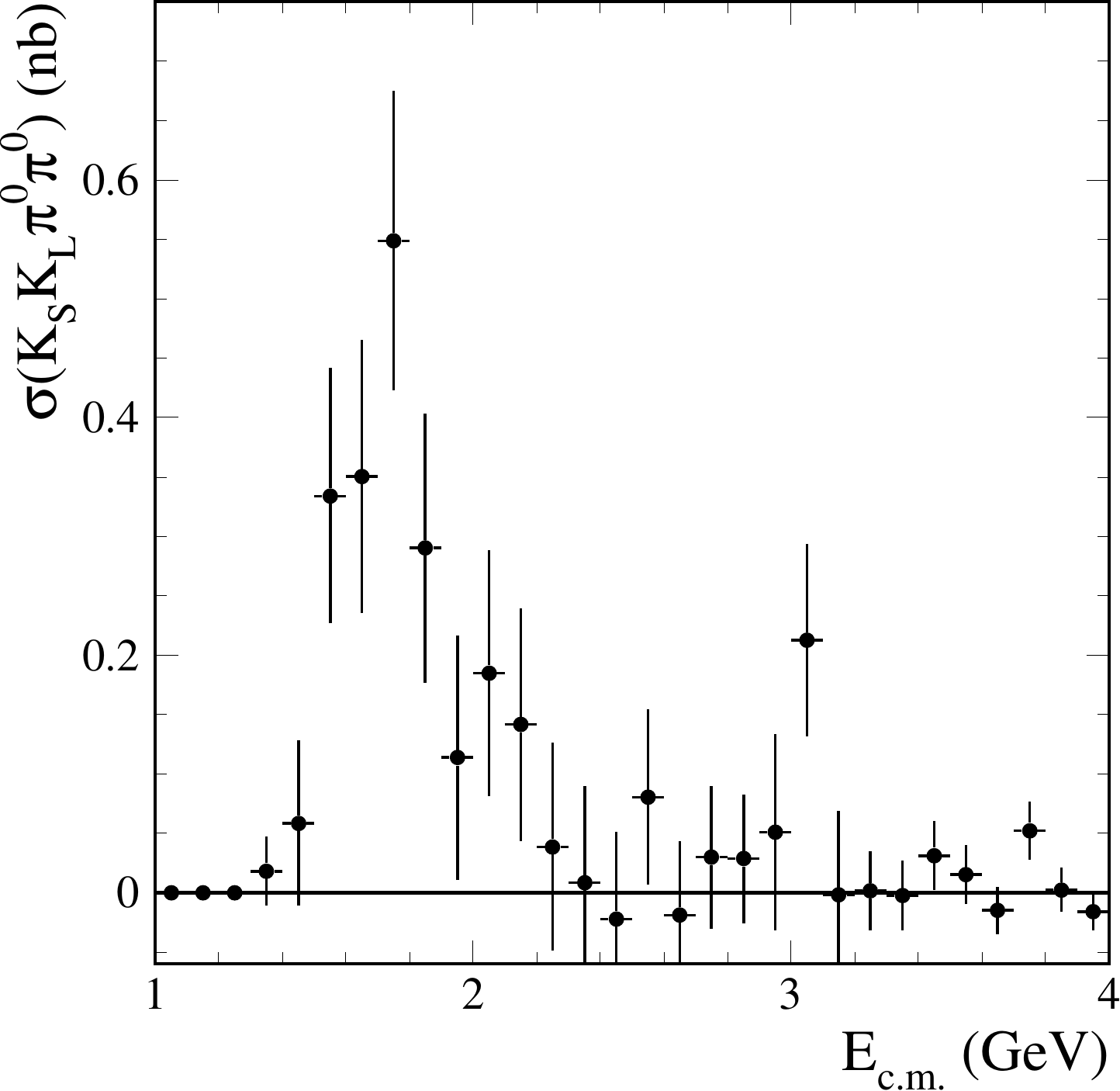}};
    \node[darkgreen,align=center] at (5.0,6) {\babar\\ preliminary \\
      \small stat. uncertainties only};
    \end{tikzpicture}
  
  \caption{Measured cross section for $\epem \to \KSKLpizpiz$.  The
    uncertainties shown are statistical only. }
  \label{fig:KSKLpizpiz-xsect}
\end{figure}

The cross section for $\epem \to \KSKLpizpiz$ is measured in an
analogous fashion.  For this channel, the region $\chi^2(\KSKLpizpiz)
< 30$ is taken as the signal region, and the control region is taken
as $30 < \chi^2(\KSKLpizpiz) < 60$.  Dominant backgrounds are from the
ISR processes $\epem \to \KS\KL$, $\KSKLpiz$, and $\KS\KL\eta$; they
are subtracted using simulated data normalised to our measurements.
Additional backgrounds are estimated from the $\chi^2$ control
region.  The resulting cross section is shown in
Fig.~\ref{fig:KSKLpizpiz-xsect}.   It rises from threshold at
$1.4\gev$ to a maximum of about $0.5\;\text{nb}$ near $1.8\gev$ and
then decreases with increasing CM energy, except for
a \jpsi and (possibly) $\psi(2S)$ signal.

The relative systematic uncertainty is dominated by the background
subtraction; it ranges from $25\%$ in the peak region of the cross
section to $60\%$ around 2\gevcc and $100\%$ at higher CM energies.

The statistics is too low to investigate the resonant substructure in
detail.  A contribution from $\Kstar(892)^0 \Kbar\piz + c.c$ is evident,
but there is no indication for $\Kstar(892)^0\Kstarb(892)^0$.  This is
consistent with the expectation from a study of the final state
$\Kp\Km\pip\pim$~\cite{Lees:2011zi}.

\subsection{Charmonium Decays to $\KSKLpiz(\piz)$}

\begin{table*}[t]
  \centering
    \begin{tabular}{llll}
      \toprule
      & \multicolumn{3}{c}{$\calB / 10^{-3}$} \\
      & \multicolumn{1}{c}{BABAR prelim.} 
      & \multicolumn{2}{c}{PDG 2014} \\
      \midrule
      $\jpsi \to \KS\KL\piz$
      & $2.06 \pm 0.24 \pm 0.10$ 
      & \\
      $\jpsi \to \KS\KL\piz\piz$
      & $1.86 \pm 0.43 \pm 0.10$ 
      & $2.35 \pm 0.41$ &(from $\Kp \Km \piz\piz$) \\
      \midrule
      $\psi(2S)  \to \KS\KL\piz$
      & $< 0.3 $
      & \multicolumn{1}{c}{---} & \\
      $\psi(2S) \to \KS\KL\piz\piz$
      & $1.24 \pm 0.54 \pm 0.06 $
      & \multicolumn{1}{c}{---} & \\
      \bottomrule
    \end{tabular}
  \caption{Branching fractions of the charmonium resonances \jpsi and
    $\psi(2S)$ decaying to the final states observed in this
    analysis.  For the decay $\jpsi \to \KSKLpizpiz$ only, there is an estimate
    of the branching fraction in the 2014 edition of the Review of
    Particle Physics \cite{Agashe:2014kda}, which was inferred from
    $\calB(\jpsi \to \Kp\Km\piz\piz)$ .}
  \label{tab:charmonium}
\end{table*}

A clear signal of $\jpsi \to \KSKLpiz(\piz)$ is visible in the cross
sections (Figs.~\ref{fig:KSKLpiz-xsect} and
\ref{fig:KSKLpizpiz-xsect}).  
Furthermore, there is some indication of the $\psi(2S)$
decaying to $\KSKLpizpiz$ but not to $\KSKLpiz$.  Using the PDG values
for $\Gamma_{ee}(\jpsi) = 5.55\kev$ and $\Gamma_{ee}(\psi(2S)) =
2.35\kev$ \cite{Agashe:2014kda}, we obtain the branching fractions 
shown in Table~\ref{tab:charmonium}.   The systematic uncertainty of
$5\%$ comes from the correction of the MC-derived efficiencies to match
data, and from variations in the procedure to determine the signal
yield.

Our measurements are the first observations of these $\jpsi$ decay modes.  The
process $\psi(2S) \to \KSKLpizpiz$ is seen with a significance of
about two standard deviations; this decay has not been reported previously.

\section{Summary}

We present studies of the processes $\epem \to \KSKLpiz$ and $\epem
\to \KSKLpizpiz$ at center-of-mass energies below 4\gev, using the
technique of initial-state radiation on a data sample of $469\invfb$
collected with the \babar detector.  The cross sections for both
processes are measured for the first time over the full energy range
from threshold to 4\gev, and the resonant sub-structure is
investigated.  We observe the decays $\jpsi \to \KSKLpiz$ and $\jpsi
\to \KSKLpizpiz$ for the first time and see evidence for $\psi(2S) \to
\KSKLpizpiz$ at a significance of 2$\sigma$.

These results complete the study of the reactions $\epem \to K
\Kbar\pi\pi$ by \babar in all possible charge combinations (except the
final states containing $\KL\KL$).  This will allow a much more precise
calculation of the contribution of the $K\Kbar\pi\pi$ final states
to the hadronic vacuum polarisation without the need to rely on
isospin relations.

\section{Acknowledgements}

We would like to thank the organizers for this very stimulating
conference. We are grateful for the excellent luminosity
and machine conditions provided by our PEPII
colleagues, and for the substantial dedicated effort
from the computing organizations that support BABAR.
The collaborating institutions wish to thank SLAC for
its support and kind hospitality. This work is supported
by DOE and NSF (USA), NSERC (Canada), CEA and
CNRS-IN2P3 (France), BMBF and DFG (Germany),
INFN (Italy), FOM (The Netherlands), NFR (Norway),
MES (Russia), MICIIN (Spain), STFC (United Kingdom).
Individuals have received support from the DFG
(Germany).


\bibliographystyle{elsarticle-numcoll}
\bibliography{KKpiz}

\begin{thebibliography}{10}
\expandafter\ifx\csname url\endcsname\relax
  \def\url#1{\texttt{#1}}\fi
\expandafter\ifx\csname urlprefix\endcsname\relax\def\urlprefix{URL }\fi
\expandafter\ifx\csname href\endcsname\relax
  \def\href#1#2{#2} \def\path#1{#1}\fi

\bibitem{Davier:2010nc}
M.~Davier, A.~Hoecker, B.~Malaescu, Z.~Zhang, {Reevaluation of the Hadronic
  Contributions to the Muon g-2 and to alpha(MZ)}, Eur. Phys. J. C71 (2011)
  1515.

\bibitem{Hagiwara:2011af}
K.~Hagiwara, R.~Liao, A.~D. Martin, D.~Nomura, T.~Teubner, {$(g-2)_\mu$ and
  $\alpha(M_Z^2)$ re-evaluated using new precise data}, J. Phys. G38 (2011)
  085003.

\bibitem{Bennett:2006fi}
G.~W. Bennett, et~al., [Muon g-2 Collaboration], {Final Report of the Muon E821
  Anomalous Magnetic Moment Measurement at BNL}, Phys. Rev. D73 (2006) 072003.

\bibitem{Grange:2015eea}
J.~Grange, [Muon g-2 Collaboration], {The New Muon g-2 Experiment at Fermilab},
  PoS NUFACT2014 (2015) 099.

\bibitem{Otani:2015jra}
M.~Otani, [E34 Collaboration], {Status of the Muon g-2/EDM Experiment at J-PARC
  (E34)}, JPS Conf. Proc. 8 (2015) 025008.

\bibitem{Redmer:Tau2016}
C.~F. Redmer, {The two-photon physics program at BESIII}, these proceedings.

\bibitem{Griessinger:Tau2016}
K.~Griessinger, {New ISR Cross Section Results on $\epem \to \pip\pim\piz\piz$
  and $\epem \to \pip\pim\eta$ from BABAR}, these proceedings.

\bibitem{Aubert:2001tu}
B.~Aubert, et~al., [BABAR Collaboration], {The \babar detector}, Nucl. Instrum.
  Meth. A479 (2002) 1--116.

\bibitem{TheBABAR:2013jta}
B.~Aubert, et~al., [BABAR Collaboration], {The \babar Detector: Upgrades,
  Operation and Performance}, Nucl. Instrum. Meth. A729 (2013) 615--701.

\bibitem{Ivanchenko:KinFit}
V.~N. Ivanchenko, private communication.

\bibitem{Aulchenko:1998xy}
V.~M. Aulchenko, et~al., [SND Collaboration], {First observation of $\phi(1020)
  \to \pi^0 \pi^0 \gamma$ decay}, Phys. Lett. B440 (1998) 442--448.

\bibitem{Aubert:2007ym}
B.~Aubert, et~al., [BABAR Collaboration], {Measurements of $e^{+} e^{-} \to
  K^{+} K^{-} \eta$, $K^{+} K^{-} \pi^0$ and $K^0_{s} K^\pm \pi^\mp$ cross-
  sections using initial state radiation events}, Phys. Rev. D77 (2008) 092002.

\bibitem{Lees:2011zi}
J.~P. Lees, et~al., [BABAR Collaboration], {Cross Sections for the Reactions
  $e^+e^- \to K^+ K^- \pi^+\pi^-$, $K^+ K^- \pi^0\pi^0$, and $K^+ K^- K^+ K^-$
  Measured Using Initial-State Radiation Events}, Phys. Rev. D86 (2012) 012008.

\bibitem{Agashe:2014kda}
K.~A. Olive, et~al., [Particle Data Group Collaboration], {Review of Particle
  Physics}, Chin. Phys. C38 (2014) 090001.

\end{thebibliography}

\end{document}